\documentclass{KapProc} 
\usepackage{amsmath,amssymb,graphicx}

\let\footnote\savefootnote

\let\footnotetext\savefootnotetext 

\let\footnoterule\savefootnoterule 

\normallatexbib

\begin{document}

\articletitle[Discrete Breathers]
{Aspects of DISCRETE BREATHERS and New Directions}

\bigskip
{\small Proceedings of NATO Advanced Research Workshop

``Nonlinearity and Disorder: Theory and Applications"

Tashkent,Uzbekistan, October, 1-6 2000

}

\chaptitlerunninghead{}

\author{S. Aubry and G. Kopidakis }
\affil{Laboratoire L\'eon Brillouin \\ 
CEA Saclay, F-91191 Gif-sur-Yvette Cedex, France }
\email{aubry@llb.saclay.cea.fr}

%%%%%%%%%%%%%%%%%%%%%%%%%%%%%%%
\begin{abstract}
We describe results concerning the existence proofs of Discrete Breathers 
(DBs) in the two classes of dynamical systems with optical linear phonons 
and with acoustic linear phonons. A standard approach is by continuation of 
DBs from an anticontinuous limit. A new approach, which is purely variational, 
is presented. We also review some numerical results on intraband DBs in random 
nonlinear systems. Some non-conventional physical applications of DBs are 
suggested. One of them is understanding slow relaxation properties of glassy 
materials. Another one concerns energy focusing and transport in biomolecules 
by targeted energy transfer of DBs. A similar theory could be used for
describing targeted charge transfer of nonlinear electrons (polarons) and,
more generally, for targeted transfer of several excitations (e.g. Davydov 
soliton).
\end{abstract}


%%%%%%%%%%%%%%%%%%%%%%%%%%%%%%%
\section{Introduction}

Discrete Breathers (DBs), also called Intrinsic Localized Modes, are 
spatially localized time-periodic solutions of discrete  classical 
nonlinear Hamiltonian systems
\footnote{We do not consider in this paper DBs in dissipative systems
which behave very differently and, in particular, can be dynamical
attractors.}. 
This self-localization is the consequence of model nonlinearity
and discreteness independent of possible disorder. DBs 
are nonlinear modes which are rather universal
and can be found in many finite or infinite 
systems of arbitrary dimension, which could be spatially periodic,
random, or else, and highly complex.
Their detailed properties may be quite diverse and model dependent
but up to now their investigation was mostly restricted to simple toy models.
DBs, which are expected to exist in complex molecules or materials,
could be studied through {\it ab initio} calculations.

Actually, their discovery has been predated by many early empirical
works during the whole last century. For example, 
the concept of \textit{local modes} was already introduced in chemistry 
for describing  localized nonlinear vibrations in molecules long ago 
\cite{BS26}, was then forgotten and rediscovered  \cite{OE82, Sco99}.
Polarons are also nonlinear objects belonging to the family of DBs 
introduced long ago by Landau \cite{Lan33}. In that case,
the coupling of a quantum electron with a deformable medium 
may induce electronic self-localization by  generating nonlinearities in the 
effective Schr\"odinger equation. 

From the mathematical point of view, breathers were also found 
in some integrable  models (Sine-Gordon) \cite{Lam82}. However, 
the breather solutions in these particular non-discrete models  have the
severe flaw to be non-generic because they disappear (as exact solutions) under
most weak model perturbations.
More recently, the discrete solitons of self-trapping equations,
including Discrete Nonlinear Schr\"odinger Equations (DNLS),
were intensively studied \cite{Sco99}.
They could be viewed under many respects (but not all) as good prototypes 
of DBs. However, they have the peculiarity to be strictly
monochromatic with only one frequency and no harmonics. 

The first claim for the generic existence of
DBs in nonlinear crystals was given by Sievers and Takeno (1988)
\cite{ST88}. However, it was still believed by many physicists that these
solutions were only approximate and it was only in 1994 that their
existence was rigorously established in infinite classical systems of
coupled nonlinear oscillators \cite{MA94}. 

The key mathematical property which makes DBs highly interesting in physics
is that they are universal and correspond to
\textit{Exact, Robust and Linearly stable} solutions of classical
nonlinear systems independent of their complexity.
Classical DBs come as one parameter family parameterized by their frequency.
Because of their universality, their analytic form can be rarely explicited,
but they can be easily calculated numerically
at any required accuracy in any given model where they exist \cite{MA96}. 
Moreover, they are robust solutions, which means that they are not specific 
to Hamiltonians with a particular form. They persist
as exact solutions of continuous families of models 
involving arbitrary potentials.
They are linearly stable which means that any small perturbation of their
initial conditions does not grow in time when treated linearly. This property
does not imply the strict stability over very long time
but it guarantees that any small initial perturbation will grow
much slower than exponentially in time. The linear stability 
property implies that when DBs are created in a nonlinear system on top of a 
quasi-linear thermalized background, they can
persist over very long times that diverge when the temperature
of the background becomes low. 

In this short proceedings paper, we shall focus only on some aspects of DBs.
We first recall the existence proof of DBs by continuation
from an anticontinuous limit and then present some details
concerning new existence criteria for hard DBs obtained by 
variational methods. We also discuss the existence of intraband DBs 
in systems that are both nonlinear and random with localized linear modes.
We also briefly discuss the spontaneous formation of 
DBs in some out of equilibrium situations and also
some exciting properties of DBs which
may be targeted selectively from donor sites to acceptor sites 
in appropriate conditions. 

\section{Existence Proofs for Discrete Breathers}
\begin{figure}[tbp]
    \centering
    \includegraphics[height=6cm]{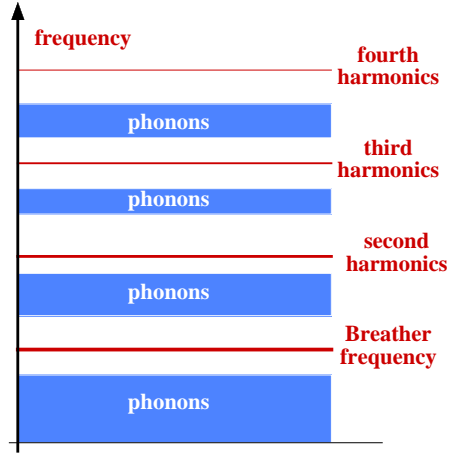}
    \caption{(Scheme) When the fundamental frequency of a DB and all its 
    harmonics lye in the gaps of the linear phonon spectrum, no radiation 
    is possible and the DB solution persists with an infinite lifetime
    (at zero temperature).}
    \label{fig1}
\end{figure}

The proof of existence of DBs essentially requires two ingredients: 
\begin{itemize}
    \item  First, the system is truly nonlinear, i.e., the frequency of
a mode depends on its amplitude (or, equivalently, its action).
\item  Second, the system is discrete 
so that the linear phonon spectrum exhibits gaps and does not extend
up to infinite frequencies.
\end{itemize}
  With these conditions, local modes may be found
with frequency and its harmonics outside the linear phonon spectrum
(see Fig.\ref{fig1}). Then, it can be understood  intuitively that 
since this local mode cannot
emit  any radiation by linear phonons (at least to leading order),
the local mode energy remains trapped. This local mode
may persist over very long time as a quasi-steady solution. 
Actually, we prove a stronger result under these assumptions,
i.e., the existence of  exact solutions with infinite lifetime.

These arguments make clear why the exact breather solutions
in the Sine-Gordon model are non-generic. There are always 
breather harmonics in the  
linear phonon spectrum which extends up to infinity and the absence 
of linear radiation can be viewed as a highly exceptional phenomenon.
Although DNLS equation is discrete, it is also particular,
since for ensuring the existence of an exact solution it suffices that the 
fundamental frequency of the DBs (there are no harmonics) does not belong to
the linear phonon spectrum.

The same arguments also suggest that the existence of quasiperiodic DBs 
\textit{in Hamiltonian systems} with extended linear phonons,
is also non-generic. The reason is that the 
time Fourier spectrum of such a solution is dense on the real axis
and necessarily overlaps with the linear phonon bands. As a result, it
should radiate energy by phonon emission till the solution either 
becomes time-periodic or vanishes.

Our initial method \cite{MA94} for proving the existence of DBs works
close enough to a  limit called anticontinuous, where the system decouples into
an array of uncoupled anharmonic oscillators and where the existence 
of local modes is trivial. Then, the implicit function theorem
can be used for proving that this exact solution persists when the 
anharmonic oscillators are coupled.

There are several ways for determining an anticontinuous limit for a 
given model. These models can be classified in two classes.
The first class involves only optical modes with a phonon gap.
The second class, which is more realistic for real materials,
involves acoustic phonons.

\paragraph{Methods using an anticontinuous limit}

Let us consider as an example of the first class, a Klein-Gordon chain 
with Hamiltonian
\begin{equation}
    H_{KG}= \sum_{i} \frac{1}{2}\dot{u}_{i}^{2}+
     V(u_{i}) +\frac{C}{2} (u_{i+1}-u_{i})^{2} 
    \label{hamKG}
\end{equation}
where the atoms $i$ with scalar coordinate $u_{i}$
and unit mass are submitted to an on-site 
potential $V(u_{i})$ with zero minimum at $u_{i}=0$
and are elastically coupled to their nearest neighbors
with coupling constant $C$. To fix the ideas, this local potential
expands for small $u_{i}$ as
\begin{equation}
    V(u_{i}) = \frac{1}{2}u_{i}^{2}+\ldots
    \label{KGpotex}
\end{equation}
Then, the linear phonon frequency at wave vector $q$ is 
\begin{equation}
    \omega^{2}(q)=\sqrt{1+4C \sin^{2} \frac{q}{2}}
\end{equation}
which yields that the phonon spectrum is the interval
$[1,\sqrt{1+4C}]$ and exhibits a gap $[0,1]$. 

An anticontinuous limit is obtained for this model at $C=0$,
when the anharmonic oscillators are uncoupled.
Then, the motion of each oscillator is periodic and its frequency 
$\omega(I)$ depends  on its amplitude or equivalently on its action 
$I$  (which is the area of the closed loop in the phase 
space $(u,\dot{u})$). Thus, we generally have $d\omega(I)/dI \neq 0$.   

There are trivial DB solutions at the anticontinuous limit
corresponding, for example, to a single oscillator oscillating at frequency 
$\omega_{b}$ while the other oscillators are immobile.
If $n\omega_{b} \neq 1$ for any integer $n$, i.e., the breather
frequency and its harmonics are
not equal to the degenerate phonon frequency $\omega(q)$ at $C=0$,
the implicit function theorem can be used for proving that this solution persists
up to some non-zero coupling $C$ as a DB solution at frequency 
$\omega_{b}$ \cite{MA94}.

More generally, even in complex models with an anticontinous limit, the 
existence of DBs can be easily proven not too far from this limit
\cite{MS95,Aub97,SM97}.

However, models of the second class with acoustic phonons create
problems because the phonon spectrum contains the frequency $0$.
Then, $n\omega_{b}$ always belongs to this phonon spectrum for
$n=0$. However, it was shown  \cite{Aub98} that in 
molecular crystal models with harmonic acoustic phonons and non-vanishing
sound velocity, the  resonant coupling between the DB
harmonics at $n=0$ and the acoustic phonons becomes harmless
and that the DB persists in the coupled system.

If one denotes as $\mathbf{y}_{i}$, the coordinates describing 
molecule $i \in \mathcal{Z}^d$ on a $d$ dimensional lattice
in its center of mass, $H_{O}(\mathbf{y}_{i})$ its 
Hamiltonian, and $\mathbf{x}_{i}$, the coordinates of its center of 
mass, the Hamiltonian of such systems is assumed to
have the form
\begin{eqnarray}
	H_{Mol} &=& \sum_{i} H_{O}(\mathbf{y}_{i}) +
	k^{\prime} \sum_{<i,j>} W(\mathbf{y}_{i},\mathbf{y}_{j})
	\nonumber \\
 && +\frac{1}{2} \mathbf{X}^t.\mathbf{N}. \mathbf{X} 
 +  \frac{1}{2} \mathbf{\dot{X}}^t.\mathbf{M}.\mathbf{\dot{X}} 
 + k \mathbf{Y}^t.\mathbf{P}.\mathbf{X} 
\label{hammol}
\end{eqnarray}
where $<i,j>$ denotes nearest neighbor molecules.
In this expression, $\mathbf{X}=\{\mathbf{x}_{i}\}$ represents the set of
variables called acoustic and $\mathbf{Y}=\{\mathbf{y}_{i}\}$
represents the set of internal variables of the molecules called optical.
$k^{\prime} W(\mathbf{y}_{i},\mathbf{y}_{j})$ represents a small direct 
nonlinear coupling between the optical variables of nearest neighbor molecules.
The elastic energy of the crystal is assumed to
be quadratic and characterized by the translationally invariant matrix
$\mathbf{N}$. $\mathbf{M}$ is a diagonal matrix with 
the molecule mass $M$ as constant diagonal.
There is also an elastic coupling energy with matrix $k \mathbf{P}$
between the acoustic variables
$\mathbf{X}$ and the optical variables $\mathbf{Y}$
obeying to translation invariance as well.

One gets an anticontinuous limit when coefficients $k$ and
$k^{\prime}$ are zero, i.e., the molecules are uncoupled.
Then, the dynamics of acoustic variables uncoupled from the
optical variables exhibit a standard gapless phonon spectrum 
with sound velocities generally depending on the direction in the 
crystal.

Then, it is proven \cite{Aub98} that if the uncoupled molecules exhibit
a nonlinear mode (i.e., a time-periodic solution) with frequency
and harmonics $n\omega_{b}$ that do not
belong to the acoustic phonon spectrum for any non-zero integer $n$,
and are not equal to any of the linear (i.e. normal)  modes of the isolated 
molecule (for any $n$), then the coupled system with $(k,k^{\prime}) \neq 0$ 
not too large
does exhibit DB solutions. Only generic assumptions are necessary for 
this result (non vanishing-sound velocities and dependence of the frequency
of the nonlinear mode of the molecule on the action).

\begin{figure}[tbp]
    \centering
    \includegraphics[height=5cm]{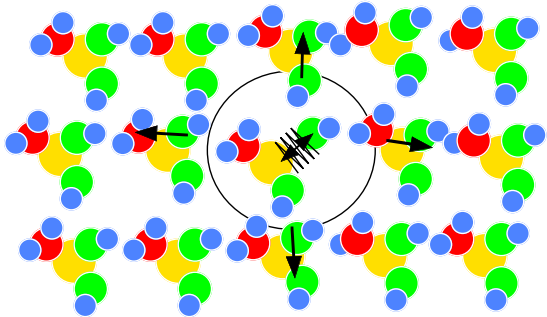}
    \caption{The nonlinear mode of an isolated anharmonic molecule may persist 
    in the crystal as a DB  when there are no resonances between its frequency 
    (and harmonics) and the linear phonon modes of the whole system.}
    \label{fig2}
\end{figure}

These DBs are obtained by continuation with respect to $k$ and $k^{\prime}$ of
the time-periodic solution at $(k,k^{\prime}) =0 $ where a single molecule oscillates
(see Fig.\ref{fig2}). Their frequencies may belong to the gap(s) between the 
linear acoustic modes and the linear optical modes or may be above the 
whole phonon spectrum. However, it is important to note that while 
the dynamical part of the DB decreases exponentially,
the DB behaves like a static impurity, i.e.,
there is generally a static component which decays with a power 
law (two and more dimensions). 

Another technique for proving the existence of DBs in diatomic FPU models was used 
in \cite{LSM97}. The Hamiltonian of such models has the form
\begin{equation}
    H_{FPU}= \sum_{i} \frac{1}{2} m_{i} \dot{u}_{i}^{2}+
     W(u_{i+1}-u_{i}) 
    \label{hamFPU}
\end{equation}
where $m_{i}=m$ for $i$ even and $m_{i}=M$ for $i$ odd.
$W(x)$ is an anharmonic convex potential.
An anticontinuous limit for this model is obtained 
for a zero mass ratio $m/M$.
The dynamical equation of this system is
\begin{equation}
   m_{i} \ddot{u}_{i} + 
   W^{\prime}(u_{i}-u_{i-1})-W^{\prime}(u_{i+1}-u_{i}) =0
    \label{dyneq}
\end{equation}
When the mass of the heavy odd atoms $M$ goes to infinity, a solution of 
these equations is obtained for $u_{2i+1}$ time-independent
($\ddot{u}_{2i+1}=0$). The dynamical equations for the light even atoms 
decouple
\begin{equation}
   m \ddot{u}_{2i} + 
   W^{\prime}(u_{2i}-u_{2i-1})-W^{\prime}(u_{2i+1}-u_{2i}) =0
    \label{dyneq2}
\end{equation}
so that the motion of the even atoms $2i$ are those of 
uncoupled anharmonic oscillators with fixed potential 
$W(u_{2i}-u_{2i-1})+W(u_{2i+1}-u_{2i})$.
The simplest DB at this limit corresponds to a single even atom oscillating
in the chain while the heavy atoms are relaxed at static equilibrium.
Then, under standard non-resonance conditions (and omitting technical details),
the implicit function theorem also states in this case,
that this time-periodic solution 
persists when $M$ is not infinite but only large enough.
However, this approach cannot prove 
the existence of a DB solution when the mass ratio becomes unity 
($M=m$).

\paragraph{Variational methods}
A flaw of the methods proceeding by continuation of DBs from an anticontinuous limit is that 
it is generally difficult to estimate explicitly the domain of parameters where 
DBs persist (although this can be easily tested numerically).  This is the reason
why we started a new approach  
based on variational methods  \cite{AKK01} which yields criteria 
ensuring the existence of DBs for a given model at given parameters.
This variational approach is inspired from the existence criteria 
for polarons (which can also be used for DNLS models).
In this simpler case, it suffices to exhibit a wavefunction with variational 
energy smaller than the bottom of the electronic band.

Up to now, we found such a criterion for some classes of models
proving only the existence of hard DBs (i.e., with frequency above the 
linear phonon spectrum). As an application, this criterion can be used 
for the above FPU diatomic chain when the potential $W(x)$ is hard (i.e., when
$W(x)$ diverges faster than $x^{2}$ for large $x$).  Then, there are DBs
at large enough frequencies. Moreover, in special cases, like the 1D
$\beta$-FPU model where $W(x)=x^{2}/4+x^{4}/4$, 
it can be proven with our criterion that there are DBs with 
frequencies infinitely close to the acoustic phonon band edge.
Note that this result is not valid for $\beta$-FPU models in two and more 
dimensions, where it is known that there is an energy threshold for DBs 
\cite{FKM97}.
Our criterion can be used (or easily extended) for complex Hamiltonians 
on periodic lattices with massive particles with hard DBs.

This criterion applies to translationally invariant Hamiltonians describing atoms
on a periodic lattice, e.g. $\mathcal{Z}^{d}$, submitted to a potential
with local interactions which is convex at least in some vicinity 
of its ground-state $u_{i} \equiv 0$.  In order to fix the ideas,
we consider for example the form
\begin{equation}
   H = \sum_{i} \frac{1}{2} \dot{u}_{i}^{2} + V(u_{i}) +
   \sum_{<i,j>} W(u_{j}-u_{i})
    \label{hammass}
\end{equation}
where $u_{i}$ are scalar variables describing the displacements of 
particles with mass unity (taken equal for simplicity) and where $V(x)$ and $W(x)$ are
anharmonic convex functions. We can set $V(x)\equiv 0$ for obtaining 
a model of the second class with acoustic phonons. Despite the fact that we have
to construct different proofs for the two cases, the final criterion for
DB existence is the same. Then, Klein-Gordon chains and FPU 
models correspond to special cases of this model.

We  define first for $\mathcal{C}_{2}$, $2\pi$ periodic loops 
$\{u_{i}(\varphi)\}$
(i.e., twice differentiable with respect to $\varphi$ with
continuous derivatives), a quantity we call ``pseudoaction''
\begin{equation}
   J(\{u_{i}(\varphi)\}) = \frac{1}{2\pi} 
   \int_{0}^{2\pi} \sum_{i} |\frac{d u_{i}}{d \varphi} (\varphi)|^{2} d\varphi
    \label{pseudoaction}
\end{equation}
and the average energy
\begin{equation}
   E(\{u_{i}(\varphi)\}) = \frac{1}{2\pi} 
   \int_{0}^{2\pi} \left(\sum_{i} V(u_{i}(\varphi))+\sum_{<i,j>}
   W(u_{j}(\varphi)-u_{i}(\varphi)) \right) d\varphi
    \label{avenerg}
\end{equation}

It is straightforward to prove that the extrema of the average energy
$E(\{u_{i}(\varphi)\})$ in the  space of loops at fixed 
pseudoaction $J(\{u_{i}(\varphi)\})$ are invariant loops for the 
dynamics associated with Hamiltonian (\ref{hammass}). More precisely,
if $\{u_{i}(\varphi)\}$ is such an extremum then for
\begin{equation}
  \omega_{b}^{2}=\frac{\int_{0}^{2\pi} \sum_{i,j} \frac{du_{i}}{d\varphi}.
    \frac{\partial^{2}V}{\partial u_{i} \partial  u_{j}}.
    \frac{du_{j}}{d\varphi} d\varphi}{
    \int_{0}^{2\pi}  \sum_{i} \left(\frac{d^{2}u_{i}}{d \varphi^{2}} 
    \right)^{2}d\varphi}
    \label{eq:freqsq}
\end{equation}
$\{u_{i}(\omega_{b}t)\}$ is a spatially localized time-periodic solution
of the system, i.e. a DB solution.
We prove the existence of extrema and thus of DBs by minimax methods when 
the following simple criterion is fulfilled:

\textit{Existence Criterion for DBs}:
Let us assume that for a given Hamiltonian,
we can exhibit one loop $\{a_{i}(\varphi)\}$ with zero average, i.e., 
such that for any $i$, 
\begin{equation}
  \int_{0}^{2\pi} a_{i}(\varphi) d\varphi =0
    \label{zeroav}
\end{equation}
and such that we have the inequality
\begin{equation}
  E(\{b_{i}+a_{i}(\varphi)\}) > \frac{1}{2} \Omega^{2} 
  J(\{a_{i}(\varphi)\})
  \label{DBcriter}
\end{equation}
for any vector $\{b_{i}\}$ (independent of $\varphi$).
Then, there exists an invariant loop
$\{u_{i}(\varphi)\}$ (a DB solution) such that 
$J(\{u_{i}(\varphi)\}) \leq J(\{a_{i}(\varphi)\})$.

For given models, when potentials $V(x)$ and $W(x)$ are 
symmetric, it is often easy to test this criterion 
by choosing sine loops $a_{i}(\varphi)
=c_{i} \cos \varphi$  because the minimum of the convex function 
$E(\{b_{i}+c_{i} \cos \varphi \})$
of $\{b_{i}\}$ is obtained for $b_{i}\equiv 0$. Then, it suffices to 
test the condition $E(\{a_{i}(\varphi)\}) > \frac{1}{2} \Omega^{2} 
  J(\{a_{i}(\varphi)\})$ by calculating explicitly the average energy
  (\ref{avenerg}) for this specific loop as a function of
  $\{c_{i}\}$. 
In models where hard DBs can be found numerically, it is generally easy to
choose a DB shape $\{c_{i}\}$ which roughly approaches the shape of the 
real DB and to test if it fulfills the DB existence criteria.
Choosing $c_{i}=0$ except for $c_{0}$ is often sufficient when $c_{0}$ is 
large enough. One can also use the approximate solution obtained
from Rotating Wave Approximation methods, for finding  trying loops.

When, the potentials are not symmetric, a simple trick consists
of replacing the initial potentials by  symmetrized 
potentials $V_{S}(x)=V_{S}(-x) \leq V(x)$ and $W_{S}(x)=W_{S}(-x) \leq W(x)$ 
for all $x$ defined by
\begin{eqnarray}\label{eq:1}
        V_{S}^{\prime\prime}(x)= \min 
        V^{\prime\prime}(x),V^{\prime\prime}(-x) \qquad 
        V_{S}^{\prime}(0)=0 \qquad V_{S}(0)=0\\
	\label{eq:2}
        W_{S}^{\prime\prime}(x)= \min 
        W^{\prime\prime}(x),W^{\prime\prime}(-x) \qquad 
        W_{S}^{\prime}(0)=0 \qquad W_{S}(0)=0
\end{eqnarray}
Then, if criterion (\ref{DBcriter}) is fulfilled 
for these new symmetric potentials, it is fulfilled for the initial 
ones 
\begin{eqnarray}
   E_{S}(\{u_{i}(\varphi)\}) &=& \frac{1}{2\pi} 
   \int_{0}^{2\pi} \left(\sum_{i} V_{S}(u_{i}(\varphi))+\sum_{<i,j>}
   W_{S}(u_{j}(\varphi)-u_{i}(\varphi)) \right) d\varphi \nonumber \\
   && \leq  E(\{u_{i}(\varphi)\}) 
    \label{avenergS}
\end{eqnarray}

It comes out very generally that if potential $V(x)$, or $W(x)$, or both, grow
faster than $x^{2}$, the DB existence is granted at large enough 
frequency above the phonon band.
However, at the present stage, we only proved that the pseudoaction of the invariant 
loop is smaller than or equal to the pseudoaction of the initial 
trying solution. The strict equality is not true in general as 
confirmed by examples with models where the potentials are 
not ``hardening" at large amplitude.  However, we expect to prove the 
equality if the potentials are ``hard enough" but this needs a correct 
definition within a general formalism.

The extension of this criterion is straightforward when the variables $u_{i}$
are not scalar. Then, we can describe realistic crystal structures with
many different atoms per unit cell and different masses.
We plan to modify our minimax techniques for finding modified criteria 
proving the existence of soft DBs (i.e., with frequencies inside the linear phonon gaps). 

It appears that the existence of DBs can be rigorously proven in 
an increasing number of classes of models independent of their 
complexity and thus the the concept of DB is rather universal.

\paragraph{Discrete breathers in random systems and Anderson modes}

When the Hamiltonian system is random (for example an array of random 
anharmonic oscillators), there is no mathematical 
difference concerning the DB existence proof from the anticontinuous limit 
when the frequency of the DB and its harmonics do not belong to the
linear phonon spectrum. These DBs are called extraband DBs.

When the linear modes are localized due to disorder (Anderson modes), 
it becomes an interesting question to understand the interplay between 
localization due to disorder and localization due to nonlinearity.
It was demonstrated \cite{KA99,KA00} that disorder and nonlinearity in the 
same system play a \textit{double game}. For some time-periodic solutions,
they cooperate for maintaining mode localization
thus generating intraband DBs.
However, there are other time-periodic solutions which are spatially
extended and which can transport energy. The reasons is that
resonances between oscillators with different frequencies can be 
restored by the nonlinearities, since their effective 
frequencies can be tuned by their amplitude.  

For understanding the subtlety of this effect and its complexity, it 
is necessary to involve time-periodic solutions which are 
not single DBs. These solutions can be easily found at the anticontinuous limit 
by choosing  an arbitrary subset of the uncoupled nonlinear oscillators (finite
or infinite) oscillating at the same frequency. It can be proven in the same way
as for the single DBs \cite{Aub97} that
when their frequency and harmonics do not belong to the linear phonon spectrum
(for periodic systems as well as for random systems) these so-called multiDB
solutions persist up to some non-vanishing coupling as exact solutions. 
Among them, there is an infinite number of linearly stable multiDB solutions \cite{Aub97}.

We investigated in refs.\cite{KA99,KA00}, on the base of a combination
of numerical and analytical arguments how the frequency 
of these exact  solutions could penetrate the phonon spectrum when it 
is discrete (for example when the model randomness is large enough).
We obtained the following generic conclusions:

\begin{itemize}
    \item Any extraband single DB or any spatially 
localized multiDB systematically disappears by bifurcating  with another multiDB
before its frequency penetrates or just when reaching the linear phonon 
band edge. This bifurcation is caused by the resonance between the DB frequency
and the linear Anderson modes with frequencies close to the band edge. 

\item  There are \textit{spatially extended} time periodic solutions
obtained by continuation of certain multiDBs while their 
frequency enters the phonon spectrum till a certain intraband frequency.
At this point, the multiDBs  also disappear by bifurcating with another 
multiDB. These multiDBs are characterized by the fact that the sites where 
the linear modes associated with the encountered linear resonances are 
located, are occupied by DBs (with appropriate  phases).
As a result, the resonance with the linear mode becomes ineffective.

\item Some of these multiDBs can be continued to the linear limit at zero amplitude
but remain extended modes. Conversely, the strict continuation
of localized linear Anderson modes  at nonzero amplitude is possible 
but it yields immediately extended (but sparse) multiDBs as soon as the
amplitude is non-zero.

\item However, each localized linear Anderson mode could be continued 
\textit{but in an approximate sense only} as a spatially localized intraband DB 
but there is a (generally) small discontinuity, each time its frequency 
crosses a linear mode frequency. Obviously, the resonance 
with the linear mode would transfer the DB energy to this mode  
thus defocusing its energy and destroying the DB. Since this set of linear frequencies is 
dense inside the phonon band, there are infinitely many gaps in the 
set of frequencies where this intraband DB could persist.  However, it 
has been proven in some models that this set of frequencies is non-void. 
It is a fat Cantor set with non-zero Lebesgue measure.
Moreover, in the limit of small DB amplitudes where nonlinearities 
becomes weak, the gap widths 
of this Cantor set tend to vanish so that this Cantor set tends to  be full.
\end{itemize}

\section{Perspectives on Discrete Breathers applications}

\paragraph{Manifestation of DBs in out of equilibrium states}
Because DB solutions are robust and linearly stable, they may 
show up spontaneously in molecular dynamics simulations in nonlinear models,
where they exist, especially in situations where they are out of equilibrium.

Such situations are obtained for example when a relatively big amount
of energy is deposited locally (for example by a photon absorption). 
Tight-binding molecular dynamics
on hydrocarbons clearly show that when vibrational energy 
is deposited locally on a proton, it remains 
localized over very long times. The frequencies of the anharmonic, localized 
and time-periodic vibrations do not correspond to normal modes, which strongly 
suggests that they are DBs \cite{KA01}. These and similar phenomena are invoked
for explaining observed anomalous infrared radiation of interstellar dust 
\cite{KA01,Pap00}.
DNLS systems with initial configurations very far from equilibrium
do not thermalize and spontaneously exhibit DBs \cite{RABT00}.
Moreover, norm conservation makes that for some initial conditions, DNLS
systems \textit{cannot} thermalize \cite{RCKG00} and induces the 
formation of DBs.

\begin{figure}[tbp]
    \centering
    \includegraphics[height=8cm]{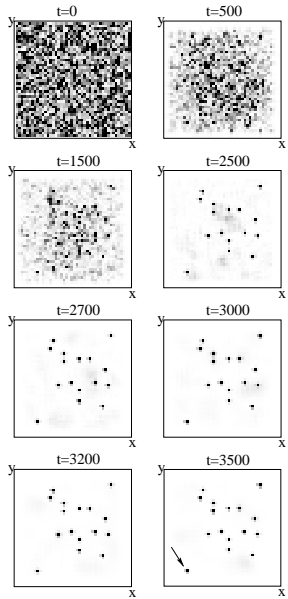}
    \caption{Time evolution of the energy density of a 
    2D nonlinear system demonstrating
    the spontaneous formation of DBs by thermal shock from \cite{BVAT99}.
    The  boundaries of the system, which is initially thermalized,
    are cooled down to zero at time zero.}
    \label{fig3}
\end{figure}

Sharp thermal shocks \cite{TA96,BVAT99} also exhibit clearly the 
formation of DBs. The example shown in Fig.\ref{fig3} was described in
ref.\cite{BVAT99}. It is obtained for a 2D array of coupled quartic oscillators
which can sustain DBs. Initially, the system is at thermal equilibrium at a 
high enough temperature in order for the energy involved in the anharmonic terms
to be relatively important. At time zero, the boundaries of the system 
are suddenly cooled down to zero K (by damping the motion of the atoms close
to the boundary). If the system were 
 harmonic, we would  expect that the total energy of the system would 
 decay exponentially as $E(t) \approx E(0) \exp (- t/\tau)$ with some 
 characteristic time $\tau$ according to the Fourier law.
Actually, the observed energy decay is much slower and can
 be well fitted by a stretched exponential $E(t) \approx E(0) \exp (- 
\alpha t^{\beta})$, where the coefficients $\alpha$ and $\beta$ 
depend on the initial temperature. If one observes the evolution of 
the spatial pattern of the energy density in the system, one finds
that after a relatively short time, it consists of narrow peaks which fit
remarkably well with exact DB solutions. It is then clear why the 
energy remains trapped in the system. In the DB regime, the energy transport 
towards the 
absorbing boundaries essentially originates from the overlap of the DBs 
tails which may generate traveling phonons from their frequency interference. 
Since this interaction becomes weaker
and weaker as the DBs density decreases and their relative distance increases, 
the rate of the energy relaxation decreases as a function of time.

This long lifetime of intraband DBs should be also useful for understanding many
anomalous energy relaxation phenomena in disordered materials (glasses, 
polymers, biopolymers).
Thermal shocks as shown above have been also performed in nonlinear random 
systems with linear Anderson modes thus implying the existence of intraband DBs
\cite{KAunp}. Again, a slow relaxation of the energy  
due to energy trapping by intraband DBs can be observed similar to Fig.\ref{fig3}.  
However, in contrast with the periodic crystal, the anomalous relaxation not 
only persists but it dramatically slows down when the initial temperature
gets small, i.e., the system becomes close to linear. This effect 
can be expected  because unlike the periodic crystal, the linear Anderson modes 
cannot propagate at all and energy propagation is essentially due to 
nonlinearities (the existence of extended multiDBs demonstrate this 
possibility).  More generally, any small thermal shock between two different 
non-zero but small temperatures reveals energy relaxation which becomes 
extremely slow when the temperature drops and no thermalization 
can be reached within computable times.
This numerical experiment suggests that there is no need to invoke static 
metastable 
configurations (usually modeled by two-level states) for interpreting the slow relaxation 
properties of real glasses since random models without 
metastable states but with only some weak nonlinearity, are sufficient 
for exhibiting a glassy behavior at low temperature. 
 
\paragraph{Targeted Energy Transfer}
Another very interesting property of DBs is that in some cases they 
could transport energy between donor and acceptor sites in a highly 
selective way. The extended dimer model described in \cite{AKMT01}
can be used for describing the transfer of energy as DBs for lattice vibrations,
electrons (polarons), excitons or many other kinds of 
excitations. It is thus highly relevant for chemical reactions 
and particularly bioenergetics which involves such transfers between different 
atoms, molecules, or macromolecules.

Let us briefly sketch the mechanism of Targeted Energy Transfer (TET)
between two anharmonic oscillators.

Resonance is the basic principle for energy transport in harmonic systems. 
The simplest example in textbooks is obtained for two weakly coupled identical harmonic oscillators.
Then, any amount of energy deposited on one of the two oscillators slowly tunnels back and forth 
between these two oscillators with a frequency proportional to the coupling.

When the oscillators are anharmonic, their frequency depends on its amplitude.
Therefore when starting the energy transfer the linear frequency  of the second (acceptor) oscillator should be 
approximately equal to the frequency of the donor oscillator oscillating at 
non-zero amplitude. 
But while the energy transfer begins the frequencies of the oscillators
change because of the change of their oscillation amplitudes and they generally
become 
unequal. Thus, the energy transfer stops much before it becomes complete.
However, there are fined tuned cases where the resonance persists 
during the energy transfer which then can become complete.
We found an analytic model for describing precisely this situation.
We call this exceptional but very important phenomenon Targeted 
Energy Transfer because it is highly selective. In 
large systems consisting of many coupled anharmonic oscillators,
it may occur only for a given amount of energy and for a selected pair of 
anharmonic oscillators called Donor 
and  Acceptor . We can also determine the critical coupling beyond which 
this energy transfer occurs.

The Donor and Acceptor oscillators are described with action-angle 
variables $I_{D},\theta_{D}$ and  $I_{A},\theta_{A}$ with Hamiltonian
$H_{D}(I_{D})$ and $H_{A}(I_{A})$. Then, the weak coupling is a 
function of $C(I_{D},\theta_{D},I_{A},\theta_{A})$ and the total 
Hamiltonian is
\begin{equation}
    H_{Dim}(I_{D},\theta_{D},I_{A},\theta_{A})=H_{D}(I_{D})+H_{A}(I_{A})
    + C(I_{D},\theta_{D},I_{A},\theta_{A})
    \label{Dimham}
\end{equation}
Since this coupling is assumed to be weak, we are close to an  adiabatic limit
where the total action is $I_{T}=2I_{0}=I_{D}+I_{A}$ is conserved and the 
half difference $I=(I_{D}-I_{A})/2$ varies slowly. The conjugate variables to 
$I_{0}$ and $I$ are  
$\theta_{0}=\theta_{D}+\theta_{A}$ and $\theta=\theta_{D}-\theta_{A}$,
respectively.
We are allowed to average  Hamiltonian (\ref{Dimham}) over $\theta_{0}$
which is a fast variable so that we obtain an integrable Hamiltonian 
with the form
\begin{equation}
    H_{T}(I_{0},I,\theta)= H_{0}(I_{0},I) +
    V(I_{0},I,\theta)
    \label{TTham} 
\end{equation}
where $I_{0}$ (which is time-constant) becomes a parameter.
$H_{0}$ is defined as
$H_{0}(I_{0},I)= H_{D}(I_{0}+I))+H_{A}(I_{0}-I)+ V_{0}(I_{0},I)$
so that the average over $\theta$ of $V(I_{0},I,\theta)$
is zero.
 
For having a solution where $I(t)$ varies from $I_{0}$ to $-I_{0}$,
thus corresponding to a \textit{complete} energy transfer
between the two anharmonic oscillators, the energy must be conserved,
which implies that unlike the harmonic case, it may occur only at specific energies
$E_{T}$ when $I_{0}$ fulfills $$H_{0}(I_{0},I_{0})=H_{0}(I_{0},-I_{0})=E_{T}$$.

With the reasonable assumption that the effective coupling $V(I_{0},I,\theta)$
is a sine-like function of $\theta$ (that is it has only one maximum 
and one minimum per period  when $|I|\leq I_{0}$),  we prove moreover that
the inequality 
\begin{equation}
  \min_{\theta} V(I_{0},I,\theta) <
  \epsilon_{T}(I)=H_{0}(I_{0},I)-H_{0}(I_{0},I_{0}) <\max_{\theta} V(I_{0},I,\theta)
  \label{condTT}
\end{equation}
is a necessary and sufficient condition for the existence of TET solution.
Function $\epsilon_{T}(I)$, defined for $|I|\leq I_{0}$ and such that
$\epsilon_{T}(I_{0})=\epsilon_{T}(-I_{0})=0$, is called 
detuning function. It may be positive, negative or change its sign in 
the interval $|I| < I_{0}$. 

\begin{figure}[tbp]
    \centering
     \includegraphics[height=4.5cm]{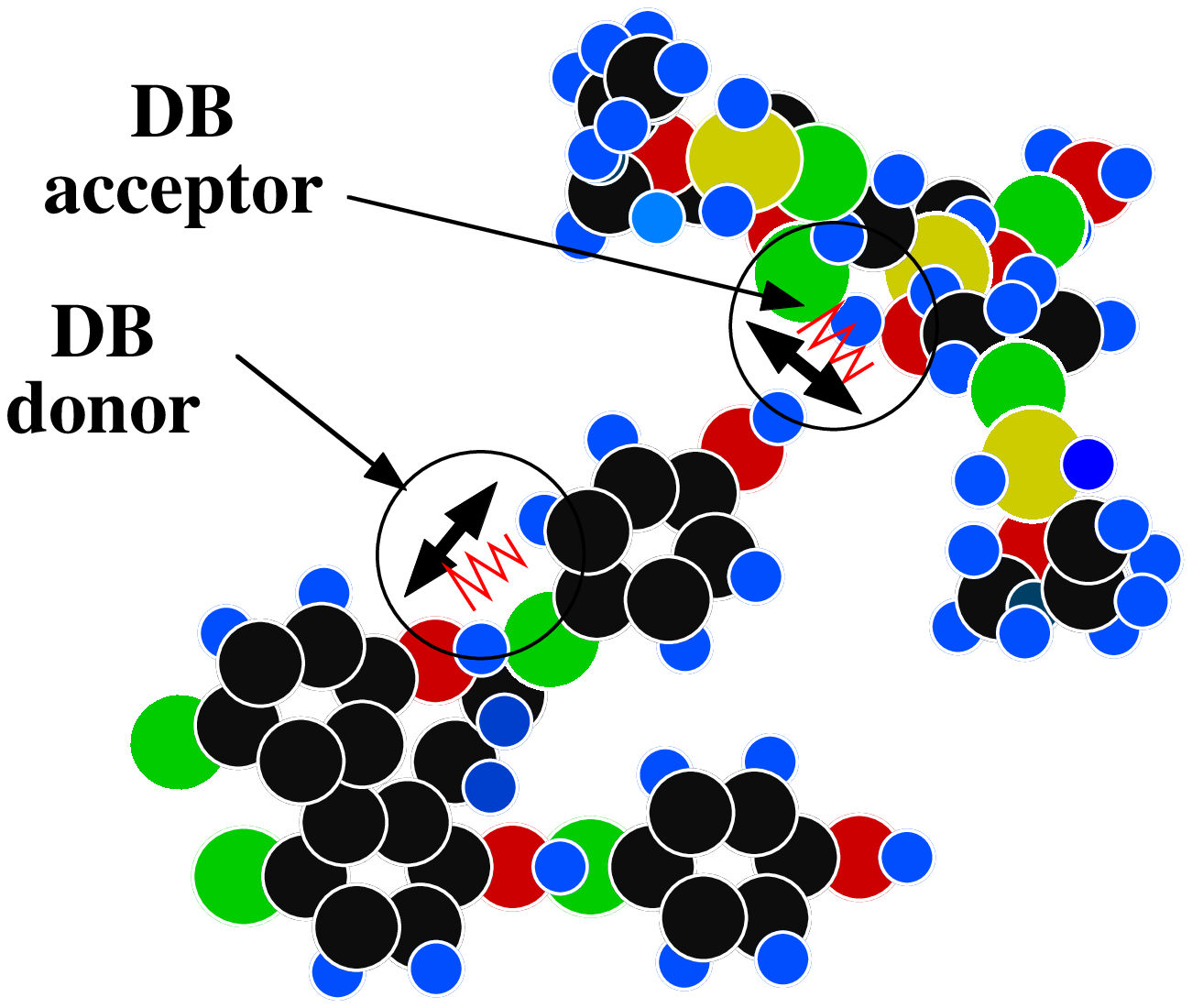}
    \includegraphics[height=3.5cm]{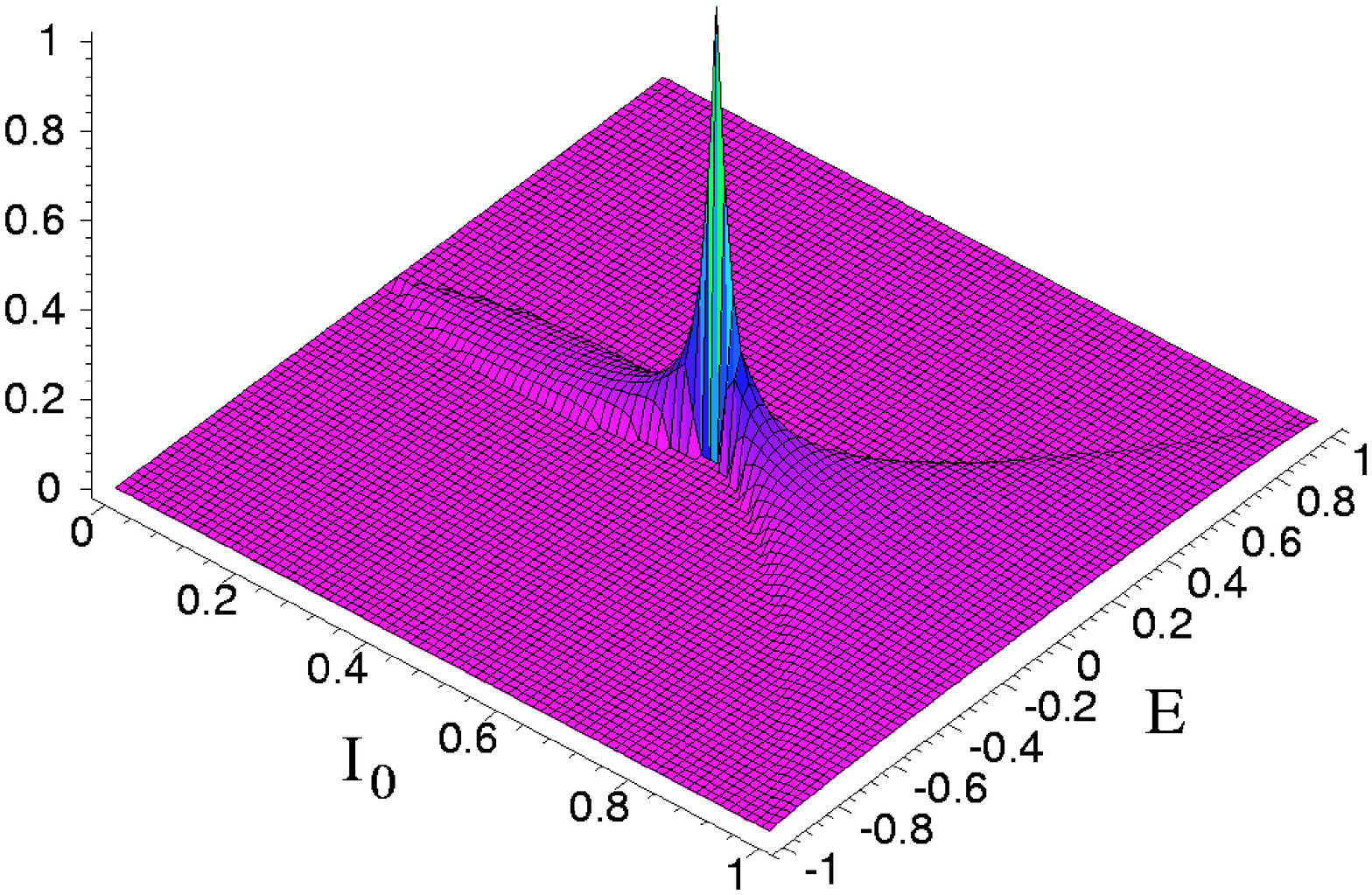}
    \caption{Schematic representation of Targeted Energy Transfer between two 
    molecules (left). Efficiency of the targeted energy transfer versus the initial energy $E$ and 
    action $I_{0}$ on the tuned dimer model. Note the very high selectivity of TET.}
    \label{fig4}
\end{figure}

Targeted Energy Transfer can be obtained at low coupling for a specific 
energy, only if the variation of the detuning function is small, which 
requires appropriately chosen Donor and Acceptor oscillators 
(particularly, one should be soft while the other is hard or vice versa).
A 3D plot of the  transfer rate between two anharmonic oscillators versus  energy $E_{T}$ and action $I_{T}$ 
is shown in Fig.\ref{fig4} for a DNLS dimer model where the
detuning function is strictly zero  \cite{AKMT01}. It demonstrates the 
high selectivity of the transfer.
 
DBs are continuous families of classical solutions which are parameterized by their frequency, or better, their
action $I$ with a total energy $H(I)$. Each DB family behaves as the periodic solutions
of an anharmonic oscillator with Hamiltonian $H(I)$ in action angle representation ($\theta$ denotes the 
variable conjugate to $I$). The DB (and multiDB) families are different if the 
systems are non-periodic. Then, in practice, conditions (\ref{condTT}) for TET can be tested numerically 
for two nonlinear systems (finite or infinite) initially 
uncoupled by calculating the energy function $H(I)$  for each pair of DBs 
families on the acceptors. These systems physically represent  
two molecules or macromolecules  Donor and Acceptor. 
Then, one can detect pairs of DBs with small detuning 
functions.  Numerical examples presented in ref.\cite{KAT01} confirm that 
for an appropriate amount of energy deposited as a specific DB on the 
donor molecule, this energy can be completely transfered as another 
specific DB on the acceptor molecule while the intermolecular coupling 
is only $2\%$  
of the intramolecular coupling.
Because of the high energy selectivity, a small energy dissipation during 
the transfer could be sufficient for making TET irreversible.
Cascades of TET along specific chains of DBs linked in a specific order
could be constructed as well as funneling of DBs toward specific centers.
We can also imagine energy stacking of DBs, etc.

The same theory can be used for describing the adiabatic transfer of a
quantum electron (or other kind of excitations such as a Davydov exciton) 
between two molecules. In this case $I_{D}$ (resp. $I_{A}$) represents 
the fraction of the electron on the Donor (resp. Acceptor) molecule 
while $H_{D}(I_{D})$ (resp. $H_{A}(I_{A})$ is the energy of the 
Donor molecule for this fractional charge (which can be defined within density 
functional theory). Thus, TET theory could be used for understanding 
chemical reactions and catalysis, especially for enzymes in biochemistry.

The new directions of applications we briefly suggested in this paper,
confirm the potential richness of the theory of Discrete 
Breathers and its  extensions. It  could become a powerful 
paradigm  for investigating a wide variety of physical phenomena in physics,
chemistry and biophysics.

\begin{acknowledgments}
This work has been supported by EC under contract HPRN-CT-1999-00163.
G. K. acknowledges support by the Greek G.S.R.T. (E$\Pi$ET II).
\end{acknowledgments}
%%%%%%%%%%%%%%%%%%%%%%%%%

\end{document}